\documentclass[]{interact}

\usepackage{epstopdf}
\usepackage[caption=false]{subfig}

\usepackage[numbers,sort&compress]{natbib}
\bibpunct[, ]{[}{]}{,}{n}{,}{,}
\renewcommand\bibfont{\fontsize{10}{12}\selectfont}
\makeatletter
\def\NAT@def@citea{\def\@citea{\NAT@separator}}
\makeatother

\theoremstyle{plain}

\theoremstyle{definition}

\theoremstyle{remark}

\begin{document}

\title{Consistency of Familial DNA Search Results in Southeast Asian Populations}

\author{
\name{Monchai Kooakachai${}^{1,*}$, Tiwakorn Chapalee${}^1$, Chairat Thitiyan${}^1$, Patsaya Jumnongwut${}^1$}
\affil{\textsuperscript{1}Department of Mathematics and Computer Science, Faculty of Science, Chulalongkorn University, Bangkok, 10330 Thailand}
}

\maketitle

\begin{abstract}
DNA databases are widely used in forensic science to identify unknown offenders. When no exact match is found, familial DNA searches can help by identifying first-degree relatives using likelihood ratios. If multiple subpopulations are relevant, likelihood ratios can be computed separately based on allele frequency estimates. Various strategies exist to combine these ratios, such as averaging allele frequencies or taking the average, maximum, or minimum likelihood ratio. While some comparisons have been made in populations like those in the U.S., their effectiveness in other regions remains unclear. This study evaluates likelihood ratio-based strategies in Southeast Asian populations, specifically Thailand, Malaysia, and Singapore. Our findings align with previous research, showing that statistical power varies across strategies. Among Thai subpopulations, the minimum likelihood ratio strategy is preferred, as it maintains high power while minimizing differences between subpopulations.
\end{abstract}


\section{Introduction}
In forensic investigation, it is very common to inquire whether a DNA sample from a crime scene matches any person in the forensic database. The problem is straightforward when an exact match is found. However, if the sample from the crime scene does not perfectly match any individual in the database but reveals some similarities, it could indicate that they are generated from two persons who are genetically related. This scenario leads to a familial DNA search, which is an approach that uses statistical tools to infer a familial relationship when a pair of DNA samples is provided. There were several attempts in developing searching strategies in many literatures, ranging from basics such as utilizing the number of alleles shared to advanced tools such as probabilistic strategies \cite{bieber2006finding, ge2011comparisons, balding2013decision, kruijver2014optimal, slooten2014probabilistic}. Familial DNA searches have gained popularity, as reported in \cite{debus2019familial}, which noted that by 2014, at least 40 laboratories in the United States had used this technique to support their investigations, with approximately 75\% of laboratories not yet using it considering future implementation.

In this context, we focus on formulating a hypothesis test to assess the relationship between two individuals or DNA profiles using a purely frequentist approach. A DNA profile consists of an individual's genotype across a specified set of loci. Each person's genotype is represented as a set of short tandem repeats (STRs), with the number of loci varying depending on the region where the analysis is conducted. Notably, the probability of two unrelated individuals having identical DNA profiles is extremely low. For instance, in a randomly mating population, the likelihood of two unrelated Caucasian Americans sharing the same 13-loci DNA profile is approximately 1 in 575 trillion \cite{reilly2001legal}.

In the analysis of familial DNA searches, the common approach, known as the likelihood ratio statistic, evaluates the ratio of the probability of encountering a pair of DNA profiles under two different relationship scenarios. One scenario represents an unrelated relationship between two persons, while another accounts for the relatedness of interest, such as a parent-child relationship. Mathematically, the likelihood ratio can be written as:
$$\dfrac{\text{P(seeing two given DNA profiles $\mid$ they are parent-child)}}{P(\text{seeing two given DNA profiles $\mid$ they are unrelated})}$$
A higher probability provides stronger evidence that the two DNA profiles are likely to belong to a parent and child. The use of the likelihood ratio statistic in kinship testing has grown steadily over the years. For example, a study on the New Zealand database \cite{curran2008effectiveness} revealed that the likelihood ratio approach for parent-child and full-sibling tests outperformed the traditional counting of matching alleles approach. Although the likelihood ratio statistic is simple and easy to interpret, it also has some limitations. One major drawback is the assumption that every individual in the population shares the same genetic background, meaning their genetic information is drawn from a single set of allele frequency distributions. However, this assumption is often inaccurate in forensic applications, where it is more realistic to account for genetic substructures within populations. Population substructure, such as racial groups or ancestry contributions, should not be ignored when calculating the likelihood ratio, as Tian et al. \cite{tian2008accounting} noted that incorporating allele frequency differences between ethnic groups can lead to distinct results in genetic studies.

One approach to incorporating population substructure, developed by the Denver Crime Lab in 2008 and inspired by \cite{presciuttini2002inferring}, involves repeatedly computing likelihood ratios using allele frequency distributions from relevant local populations. The final test statistic is determined by selecting the maximum of these computed likelihood ratios. In contrast, an alternative method suggests using the minimum likelihood ratio instead. Ge and Budowle \cite{ge2012kinship} demonstrated that the minimum likelihood ratio strategy is preferable to the maximum likelihood ratio strategy, as it tends to be closer to the true likelihood ratio. However, no evidence has yet confirmed its superiority in terms of statistical power for hypothesis testing. A 2019 study \cite{kooakachai2019new} examined likelihood ratio-based statistics for parent-child and full-sibling tests within the Colorado prison population, exploring various ways to combine likelihood ratios while accounting for population substructure. Their findings indicated that averaging local allele frequencies resulted in the highest statistical power while maintaining an acceptable false positive rate. Nevertheless, the consistency of these methods remains uncertain, as prior research has primarily focused on American populations and did not compare them with the minimum likelihood ratio approach. Investigating whether these findings hold true for Asian populations is essential, given the significant genetic differences between Asian and North American or Hispanic populations.

In this study, we evaluate the effectiveness of existing test statistics for hypothesis testing in parent-child and full-sibling cases, along with a newly proposed statistic, in terms of Type I error and power. The article is structured as follows: Section 2 provides a detailed explanation of the problem, describes the algorithm used for the simulation study, and outlines the case study. Section 3 presents the findings from the numerical analysis, emphasizing power comparisons and biases related to subpopulations. Finally, Section 4 concludes with a discussion of the implications of our findings.

\section{Methods}

Suppose a population is made up of $K$ subpopulations, each corresponding to a subgroup with a unique ethnic background, causing variations in allele frequency distributions. We denote the allele frequency distribution in subpopulation $k$ by $f_k$ where $k=1, \ldots, K.$ Also, let  $p_1, \ldots, p_K$ be the proportions of individuals in each subpopulation, such that the sum of these proportions must equal to one. 

Let $X_1$ be a DNA profile from an unknown individual observed on $m$ loci and $f_{X_1}$ be the allele frequency distribution corresponding to the subpopulation such an individual belongs to. Note that the parameter $f_{X_1}$ is generally unknown because it represents subpopulation information of the obtained forensic sample – we never know who left the sample in crime scene. Additionally, we define $X_2$ and $f_{X_2}$ for the second unknown individual in the same fashion. Table \ref{tab1} shows an example of two DNA profiles \( X_1 \) and \( X_2 \), across five independent loci: D3S1358, D7S820, D8S1179, D13S317, and D22S1045. It can be observed that \( X_1 \) and \( X_2 \) share at least one allele (identical by state) at each locus considered. This raises the question of performing a statistical hypothesis test to determine whether \( X_1 \) and \( X_2 \) have a parent-child relationship, as a father and son always share one identical by descent allele at each locus. 

\begin{table}[h!]
\centering
\caption{Example DNA profiles of two individuals, $X_1$ and $X_2$}
\begin{tabular}{|c|ccccc|}
\hline & \multicolumn{5}{c|}{Locus}  \\ \cline{2-6} 
                              & D3S1358    &  D7S820  & D8S1179     & D13S317    & D22S1045   \\ \hline
$X_1$                          & (13,14) & (11,12) & (17,18) & (13,13) & (9,10)  \\ \hline
$X_2$                          & (13,14) & (12,14) & (18,18) & (12,13) & (9,11)   \\ \hline
\end{tabular}
\label{tab1}
\end{table}

The primary goal is to develop and compare various test statistics for evaluating the following hypotheses: $H_0$, where $X_1$ and $X_2$ are unrelated individuals, versus $H_1$, where $X_1$ and $X_2$ share a specific genetic relationship.

In the alternative hypothesis, it is necessary to clarify the genetic relationship between two individuals, making it a case of simple versus simple hypotheses. The parameter here is the genealogical relatedness between two individuals, $\theta$ which can be measured by using the concept of identical by descent (IBD) alleles. In other words, any genetic relatedness between two persons can be derived as $\theta = (z_0, z_1, z_2)$ where $z_0$ is the probability that none of the four alleles from two individuals are IBD, $z_1$ is the probability that exactly one of the alleles of the first individual is IBD to one of the alleles of the second individual, and $z_2$ be the probability that both alleles of the first individual are IBD to those of the second person. For example, for a parent-child test, the hypothesis testing is $H_0: \theta = (1,0,0)$ versus $H_1: \theta = (0,1,0).$ 

Given two DNA profiles $X_1, X_2$ and corresponding allele frequency distributions $f_{X_1}, f_{X_2}$ the likelihood ratio test statistic for testing $H_0: \theta = \theta_0$ versus $H_1: \theta = \theta_1$ is defined by
\begin{equation}
\label{likelhd1}
\Lambda = \dfrac{L(\theta_1 \mid X_1,X_2,f_{X_1},f_{X_2})}{L(\theta_0 \mid X_1,X_2,f_{X_1},f_{X_2})}
\end{equation}
where $L$ represents the likelihood function for a relationship. When a population is homogeneous, the likelihood function $L$ is easily calculated because parameters $f_{X_1}$ and $f_{X_2}$ can be determined without any uncertainty. 

Table \ref{tab2} displays seven possible genotype combinations for two individuals, along with their corresponding probabilities and allele-sharing patterns for unrelated individuals, parent-child pairs, and full siblings. The formulas in each column of Table \ref{tab2} differ based on the value of $\theta$. This table is essential for computing the numerator and denominator in equation (\ref{likelhd1}). For instance, if the allele frequencies for alleles 13 and 14 at the D3S1358 locus are 0.15 and 0.20, respectively, the probability of observing the genotypes (13,14) and (13,14) under the parent-child relationship would be $(0.15)(0.20)(0.15+0.20) = 0.0105$. In contrast, assuming unrelated individuals, the probability of observing the same genotypes would be $4(0.15)^2(0.20)^2 = 0.0036.$ This results in a likelihood ratio at the D3S1358 locus of $0.0105/0.0036 \approx 2.9167.$ Such probabilities can be calculated separately for each locus and then combined by multiplication, assuming the loci are independent.

\begin{table}[h!]
\centering
\caption{Probabilities of the 7 possible combinations of genotypes for multi-allelic loci in pairs of individuals, conditional on their relationship, as functions of allele frequencies. $p_i$ represents the allele frequency of allele $i$.}
\begin{tabular}{|c|c|c|c|c|c|c|c|} 
\hline
 & Unrelated & Parent-Child & Full Sibings \\ 
Genotype & $\theta = (1,0,0)$ & $\theta = (0,1,0)$ & $\theta = (1/4,1/2,1/4)$ \\ \hline
AA,AA & $p_A^4$ & $p_A^3$ &  $p_A^2 (1+p_A)^2/4$ \\ \hline
AA,AB & $4p_A^3p_B$ & $2 p_A^2 p_B$ & $p_A^2 p_B (1+p_A)$ \\ \hline
AA,BB & $2p_A^2p_B^2$ & 0 & $p_A^2 p_B^2/2$ \\ \hline
AB,AB & $4p_A^2p_B^2$ & $p_A p_B (p_A+p_B)$ & $p_A p_B (2p_A p_B + p_A + p_B +1)/2$  \\ \hline
AA,BC & $4p_A^2p_Bp_C$ & 0  & $p_A^2 p_B p_C$ \\ \hline
AB,AC & $8p_A^2p_Bp_C$ & $2 p_A p_B p_C$ & $p_A p_B p_C (2p_A +1)$ \\ \hline
AB,CD & $8p_Ap_Bp_Cp_D$ & 0 & $2 p_A p_B p_C p_D$ \\ \hline
\end{tabular}
\label{tab2}
\end{table}

In the presence of population substructure, where multiple subpopulations exist, it is essential to determine the allele frequency distributions, denoted as $f_{X_1}$ and $f_{X_2}$. For instance, the probability of observing allele 13 at the D3S1358 locus may differ between populations—it could be 0.15 in one subpopulation and 0.18 in another. This variation creates ambiguity in selecting the appropriate value for calculations. The parameters $f_{X_1}$ and $f_{X_2}$ are called nuisance parameters as they are parameters that are not of direct interest but need to be accounted for in the analysis to avoid bias or errors in the estimation of the parameters of primary interest.

There are several existing methods for handling multiple sets of available allele frequencies. First, we can average all allele frequencies before incorporating them into the likelihood function calculation. Specifically, we define $$f_{local} = p_1f_1 + p_2f_2 + \cdots + p_Kf_K$$ and use $f_{local}$ within the likelihood ratio framework. Mathematically, this can be expressed as follows:
\begin{equation*}
\Lambda_{LAF} = \dfrac{L(\theta_1 \mid X_1,X_2,f_{local})}{L(\theta_0 \mid X_1,X_2,f_{local})}.
\end{equation*}
In this approach, both $f_{X_1}$ and $f_{X_2}$ are set equal to the local allele frequency $f_{local}$. This method, introduced in \cite{kooakachai2019new}, is referred to as LRLAF. Beyond averaging allele frequencies, alternative algebraic methods can be applied at the likelihood ratio level. Instead of averaging allele frequencies first, we can compute the likelihood ratio separately for each subpopulation and then explore different ways of combining them. One such approach involves using the population proportions as weights for each likelihood ratio, leading to the following statistic:
\begin{equation*}
\Lambda_{AVG} = p_1 \dfrac{L(\theta_1 \mid X_1,X_2,f_{1})}{L(\theta_0 \mid X_1,X_2,f_{1})} 
+ p_2 \dfrac{L(\theta_1 \mid X_1,X_2,f_{2})}{L(\theta_0 \mid X_1,X_2,f_{2})}
+ \cdots 
+ p_K \dfrac{L(\theta_1 \mid X_1,X_2,f_{K})}{L(\theta_0 \mid X_1,X_2,f_{K})} 
\end{equation*}
We denote this statistic $\Lambda_{AVG}$ as LRAVG. Additionally, instead of averaging, we can take either the maximum or minimum likelihood ratio across all available subpopulations, resulting in the following statistics (referred as LRMAX and LRMIN, respectively):
\begin{equation*}
\Lambda_{MAX} = \max \left\{ \dfrac{L(\theta_1 \mid X_1,X_2,f_{1})}{L(\theta_0 \mid X_1,X_2,f_{1})},
\dfrac{L(\theta_1 \mid X_1,X_2,f_{2})}{L(\theta_0 \mid X_1,X_2,f_{2})},
\ldots, \dfrac{L(\theta_1 \mid X_1,X_2,f_{K})}{L(\theta_0 \mid X_1,X_2,f_{K})} \right\} 
\end{equation*}
\begin{equation*}
\Lambda_{MIN} = \min \left\{ \dfrac{L(\theta_1 \mid X_1,X_2,f_{1})}{L(\theta_0 \mid X_1,X_2,f_{1})},
\dfrac{L(\theta_1 \mid X_1,X_2,f_{2})}{L(\theta_0 \mid X_1,X_2,f_{2})},
\ldots, \dfrac{L(\theta_1 \mid X_1,X_2,f_{K})}{L(\theta_0 \mid X_1,X_2,f_{K})} \right\} 
\end{equation*}
In \cite{ge2012kinship}, the authors demonstrated that taking the minimum likelihood ratio is a preferable approach to the maximum, as simulations indicated that it provides a logarithm of the likelihood distribution closest to the true values. Our goal is to validate this finding in terms of statistical power. Additionally, we introduce a new alternative approach for comparison. Instead of computing the maximum or minimum likelihood ratio directly, we can calculate the maximum or minimum likelihood functions separately under the null and alternative hypotheses and then combining them as a ratio later. These methods are defined as follows:
\begin{equation*}
\Lambda_{RMAX} =   \dfrac{ \max \left\{ L(\theta_1 \mid X_1,X_2,f_{1}), L(\theta_1 \mid X_1,X_2,f_{2}), \ldots, L(\theta_1 \mid X_1,X_2,f_{K}) \right\} }{\max \left\{ L(\theta_0 \mid X_1,X_2,f_{1}), L(\theta_0 \mid X_1,X_2,f_{2}), \ldots, L(\theta_0 \mid X_1,X_2,f_{K}) \right\}}
\end{equation*}
\begin{equation*}
\Lambda_{RMIN} =   \dfrac{ \min \left\{ L(\theta_1 \mid X_1,X_2,f_{1}), L(\theta_1 \mid X_1,X_2,f_{2}), \ldots, L(\theta_1 \mid X_1,X_2,f_{K}) \right\} }{\min \left\{ L(\theta_0 \mid X_1,X_2,f_{1}), L(\theta_0 \mid X_1,X_2,f_{2}), \ldots, L(\theta_0 \mid X_1,X_2,f_{K}) \right\}}
\end{equation*}
We refer to these two approaches as LRRMAX and LRRMIN, respectively. While LRRMAX was previously introduced in \cite{kooakachai2019new}, it has not been applied to any populations beyond the Colorado prison population. In contrast, LRRMIN is a completely new approach.

\subsection{Algorithm for Power Comparison}

To evaluate the performance of each statistic, we need to compare their power while maintaining the same false positive rate (FPR). Since there is no direct algebraic method for this comparison, we will assess power through simulation. The numerical study requires allele frequency distributions for each subpopulation $(f_1, f_2, \ldots, f_k)$, along with estimates of the proportion $p_i$ of each subpopulation. For each hypothesis test concerning the relationship of interest (parent child or full sibs), we follow the algorithm outlined below.
\begin{enumerate}
\item Simulate a pair of unrelated DNA profiles (using $m$ loci). The subpopulation information of individuals is pulled at random from a multinomial distribution. Probabilities in multinomial distribution are estimates of proportions.
\item Calculate the likelihood ratio (LR)-based statistics for a pair drawn in step 1. Then save the value. Note that the LR-based statistics considered here include LRLAF, LRAVG, LRMAX, LRMIN, LRRMAX, and LRRMIN.
\item Repeat steps 1 and 2 so many times, say $B = 1,000,000$ times. Here, we obtain the null distribution of likelihood ratios based on those $B$ values.
\item	Estimate a threshold $c$ from the null distribution in step 3. That is, find $c$ such that $\mathbb{P}(LR > c)=\alpha$ where $\alpha$ is a given false positive rate. Note that the threshold varies for each statistic.
\item Generate a pair of DNA profiles based on the relationship in the alternative hypothesis (using $m$ loci). For each pair, the subpopulation information of the first individual is selected randomly from a multinomial distribution as introduced in step 1 and the subpopulation information of the second individual is forced to be the same as the first person because they are genealogical related.
\item Calculate the likelihood ratio (LR)-based statistic for a pair drawn in step 5. Then save the value.
\item Repeat steps 5 and 6 so many times, say $B = 1,000,000$ times. Therefore, we obtain the alternative distribution of likelihood ratios based on those samples. 
\item Estimate power of each statistic by calculating the ratio of the number of pairs such that LR is greater than the threshold $c$ and the total number of simulated pairs in step 7. 
\end{enumerate}

In this study, three specific tests are focused. The first one is $H_0: \theta = (1,0,0)$ versus $H_1: \theta = (0,1,0)$, which tests unrelated pairs against parent-child relationships. The second relationship tests unrelated individuals versus full siblings and can be expressed as $H_0: \theta = (1,0,0)$ versus $H_1: \theta = (1/4,1/2,1/4)$. Lastly, the half-sibling test can be formulated as $H_0: \theta = (1,0,0)$ versus $H_1: \theta = (0,1/2,1/2)$. It is also important to note that we assume independence between loci, which allows us to multiply the probabilities computed for each locus when performing calculations on these $m$ loci.

\subsection{Case Study Description}
Three Southeast Asian populations—Thailand, Singapore, and Malaysia—were selected as a case study. Previous studies, both archaeological and linguistic, have supported the notion that Thailand has a complex genetic structure. The genetic differences among subgroups in Thailand were confirmed by Kutanan et al. (2014) \cite{kutanan2014forensic}, who revealed distinctions in genetic structure between Thai-Malay Muslims, primarily residing in the southern part of Thailand, and Thai Buddhists, who predominantly live in other regions of the country. In the same year, genetic relationships between northeastern Thai people and other subpopulations were investigated based on mitochondrial DNA-HVR1 variation, suggesting a genetic distinction between northern and northeastern Thai subgroups \cite{kutanan2014mitochondrial}. Furthermore, a genome-wide study involving 32 subpopulations in Thailand \cite{kutanan2021reconstructing} indicates the need for reconstructing the human genetic history of mainland Southeast Asia. The authors highlighted the need for careful selection of Thai populations in ongoing biomedical or clinical studies due to the significant genetic differences among various subpopulations. Considering these literature reviews, it is highly likely that Thailand exhibits population substructure. For simplicity, we assume four subpopulations in Thailand based on the main regions where Thai people reside: North, Northeastern, Central, and South. The allele frequency distributions for each subpopulation are derived from Shotivaranon et al. \cite{shotivaranon2009dna}, in which DNA was extracted from 929 unrelated Thai individuals at $m$ = 15 loci, including FGA, TH01, TPOX, CSF1PO, vWA, D2S1338, D3S1358, D5S818, D7S820, D8S1179, D13S317, D16S539, D18S51, D19S433, and D21S11. The sample sizes for the North, Northeastern, Central, and South subgroups are 202, 304, 212, and 211, respectively. For shorthand notation, we denote the allele frequency distributions for North, Northeastern, Central, and South subgroups by $f_{N}, f_{NE}, f_{CT}$ and $f_{S}$, respectively. The corresponding $f_{local}$ in this case is $p_{N}f_{N} + p_{NE}f_{NE} + p_{CT}f_{CT} + p_{S}f_{S}$ where the prior probabilities $({p}_N, p_{NE}, p_{CT},p_S)$ = (0.1108, 0.3695, 0.3538, 0.1659) estimated from Thai population census. 

The Singaporean database utilizes 15 loci spread across 23 chromosomes. Some of these loci overlap with those in the Thai database, but not all of them. In particular, the core loci include FGA, TH01, TPOX, CSF1PO, vWA, D3S1358, D5S818, D7S820, D8S1179, D13S317, D16S539, D18S51, D21S11, Penta D, and Penta E. According to a previous study \cite{yong2004allele}, the population of Singapore comprises three primary ethnic groups: Chinese (C), Malays (M), and Indians (I). The estimated proportions of these subpopulations are 77.13\% Chinese, 15.24\% Malays, and 7.63\% Indians. Lastly, the Malaysian DNA database relies on 9 loci distributed across 23 chromosomes. The key loci in this database are FGA, TH01, TPOX, CSF1PO, vWA, D3S1358, D5S818, D7S820, and D13S317. Previous research \cite{lim2001str} identifies three main ethnic groups in Malaysia: Chinese (C), Malays (M), and Indians (I), with approximate population proportions of 23.43\% Chinese, 69.49\% Malays, and 7.08\% Indians.

\section{Results}

This section presents the simulation results, organized into two key subsections: one comparing power and the other examining bias across subpopulations.

\subsection{Power Comparison}

Table \ref{tab:table3} presents a power comparison for the full-sibling test at an FPR of 0.0002 across three subpopulations. This FPR was selected for two main reasons. First, since power increases and eventually approaches 1 as the FPR rises, choosing an appropriate FPR allows for meaningful power differences to be observed; otherwise, all statistics would yield power values close to 1. Conversely, if the FPR is too low, the resulting power would also be low. Second, this FPR was chosen to align closely with values used in previous studies. In \cite{kooakachai2019new}, the authors identified 0.000012 as a reasonable FPR for the full-sibling test in the Colorado prison population, making 0.0002 a suitable compromise between maintaining comparability and ensuring detectable differences in power.

\begin{table}[h!]
\centering
\caption{\label{tab:table3} Power comparison for the full-sibling test at $\alpha = 0.0002$.}
\begin{tabular}{cccccccccc}
 &\multicolumn{3}{c}{Thai population}\\
 Statistic & Threshold & Power & 95\% CI  \\ \hline
 $\Lambda_{LAF}$ & 411.7 & 0.781 & (0.7805, 0.7821) \\
 $\Lambda_{AVG}$ & 995.1 & 0.757 & (0.7561, 0.7578)  \\
 $\Lambda_{MAX}$ & 2489.7 & 0.741 & (0.7401, 0.7419) \\
 $\Lambda_{MIN}$ & 253.8 & 0.775 & (0.7741, 0.7757)  \\
  $\Lambda_{RMAX}$ & 705.2 & 0.662 & (0.6606, 0.6625)  \\
   $\Lambda_{RMIN}$ & 89875.2 & 0.607 & (0.6060, 0.6079)  \\
\end{tabular}
\begin{tabular}{cccccccccc}
&\multicolumn{3}{c}{Singaporean population}\\
 Statistic & Threshold & Power & 95\% CI  \\ \hline
 $\Lambda_{LAF}$ &  310.5 & 0.804 & (0.8029, 0.8044)  \\
 $\Lambda_{AVG}$ & 560.1 & 0.803 & (0.8026, 0.8042)  \\
 $\Lambda_{MAX}$ & 2592.1 & 0.776 & (0.7757, 0.7773)  \\
 $\Lambda_{MIN}$ & 214.3 & 0.810 & (0.8087, 0.8103)  \\
   $\Lambda_{RMAX}$ & 648.2 & 0.656 & (0.6546, 0.6565)  \\
   $\Lambda_{RMIN}$ & 264220.7 & 0.640 & (0.6388, 0.6407)  \\
\end{tabular}
\begin{tabular}{cccccccccc}
&\multicolumn{3}{c}{Malaysian population}\\
 Statistic & Threshold & Power & 95\% CI \\ \hline
 $\Lambda_{LAF}$ & 489.7 & 0.369 & (0.3681, 0.3697) \\
 $\Lambda_{AVG}$ & 593.6 & 0.361 & (0.3605, 0.3621) \\
 $\Lambda_{MAX}$ & 1973.4 & 0.328 & (0.3268, 0.3283) \\
 $\Lambda_{MIN}$ & 331.1 & 0.357 & (0.3559, 0.3575) \\
    $\Lambda_{RMAX}$ & 399.6 & 0.272 & (0.2713, 0.2731)  \\
   $\Lambda_{RMIN}$ & 13487.0 & 0.262 & (0.2612, 0.2629)  \\
\end{tabular}
\end{table}

All thresholds shown in Table \ref{tab:table3} are referred to as null thresholds and are used for power calculations. While these values may not be particularly informative for comparing different statistics, they hold practical significance in real-world applications. For example, when using $\Lambda_{LAF}$ as the test statistic for a full-sibling test, the rejection region would be set at approximately 411.7 or higher for the Thai population and approximately 310.5 or higher for the Singaporean population. If the calculated likelihood ratio falls within this region, we would conclude that there is sufficient evidence to determine that the two individuals are full siblings.

As described in the algorithm, the power of each statistic is estimated based on the proportion of simulated pairs under the alternative hypothesis (where the relationship is assumed) whose statistic is greater than or equal to the null threshold. Additionally, the 95\% confidence interval for power is provided, calculated using the exact binomial test.

For the Thai population, $\Lambda_{LAF}$ exhibits the highest performance, with a power of approximately 78\%, followed by $\Lambda_{MIN}$, $\Lambda_{AVG}$, $\Lambda_{MAX}$, $\Lambda_{RMAX}$ and finally $\Lambda_{RMIN}$. The differences are significant, as the 95\% confidence intervals do not overlap. For the Singaporean population, $\Lambda_{MIN}$ performs the best, with a power of around 81\%, but there is no clear distinction between $\Lambda_{LAF}$ and $\Lambda_{AVG}$. In fact, the power estimates for the top three statistics in this population are quite similar, making it difficult to draw definitive conclusions. As in the Thai population, $\Lambda_{MAX}$, $\Lambda_{RMAX}$ and $\Lambda_{RMIN}$ perform the worst. In the Malaysian population, $\Lambda_{LAF}$ again leads, followed by $\Lambda_{AVG}$, $\Lambda_{MIN}$, $\Lambda_{MAX}$, $\Lambda_{RMAX}$ and  $\Lambda_{RMIN}$. Notably, $\Lambda_{MAX}$, $\Lambda_{RMAX}$ and  $\Lambda_{RMIN}$ consistently underperform compared to $\Lambda_{LAF}$, $\Lambda_{AVG}$, and $\Lambda_{MIN}$ across all three subpopulations. 

These findings are consistent with those reported in the Colorado prison population study \cite{kooakachai2019new}, where $\Lambda_{LAF}$ showed the highest statistical power, followed by $\Lambda_{AVG}$, while $\Lambda_{MAX}$ ranked the lowest. However, it is somewhat unexpected that $\Lambda_{RMAX}$ and $\Lambda_{RMIN}$ performed poorly in Southeast Asian populations, given that in the Colorado study, $\Lambda_{RMAX}$ exhibited significant power comparable to $\Lambda_{LAF}$. Further elaboration on these outcomes will be provided in the upcoming section.

Next, we take a different approach to power comparison by examining it across various relationship types. We analyze 3 relationships: parent-child, full-sibling, and half-sibling, using FPRs of 0.00002, 0.0002, and 0.002, respectively. The partial results of these comparisons are summarized in Table \ref{tab:table4}.

\begin{table}[h!]
\centering
\caption{\label{tab:table4} Power comparison across different genetic relationships.}
\begin{tabular}{cccccccccc}
 &\multicolumn{3}{c}{Parent-child at $\alpha = 0.00002$}\\
 Statistic & Thai & Singaporean & Malaysian  \\ \hline
 $\Lambda_{LAF}$ & 0.835 & 0.791& 0.167  \\
 $\Lambda_{AVG}$ & 0.816 & 0.802 & 0.153  \\
 $\Lambda_{MAX}$ & 0.800 & 0.729 & 0.116  \\
 $\Lambda_{MIN}$ & 0.823 & 0.786 & 0.148  \\
  $\Lambda_{RMAX}$ & 0.545 & 0.490 & 0.019  \\
   $\Lambda_{RMIN}$ & 0.513 & 0.423 & 0.021  \\
\end{tabular}

\begin{tabular}{cccccccccc}
 &\multicolumn{3}{c}{Full-sibling at $\alpha = 0.0002$} \\
 Statistic & Thai & Singaporean & Malaysian \\ \hline
 $\Lambda_{LAF}$ &  0.781 & 0.804 & 0.369  \\
 $\Lambda_{AVG}$ &  0.757 & 0.803 & 0.361  \\
 $\Lambda_{MAX}$ & 0.741 & 0.776 & 0.328  \\
 $\Lambda_{MIN}$ & 0.775 & 0.810 & 0.357  \\
  $\Lambda_{RMAX}$ & 0.662 & 0.656 & 0.272  \\
   $\Lambda_{RMIN}$ & 0.607 & 0.640 & 0.262  \\
\end{tabular}

\begin{tabular}{cccccccccc}
&\multicolumn{3}{c}{Half-sibling at $\alpha = 0.002$}\\
 Statistic & Thai & Singaporean & Malaysian \\ \hline
 $\Lambda_{LAF}$ & 0.984 & 0.934 & 0.607 \\
 $\Lambda_{AVG}$  & 0.973 & 0.989 & 0.632 \\
 $\Lambda_{MAX}$ & 0.948 & 0.968 & 0.548 \\
 $\Lambda_{MIN}$ & 0.986 & 0.993 & 0.636 \\
  $\Lambda_{RMAX}$ & 0.582 & 0.556 & 0.175  \\
   $\Lambda_{RMIN}$ & 0.522 & 0.554 & 0.188  \\
\end{tabular}
\end{table}

A clear pattern that emerges is that the performance of each statistic can be grouped into three main categories. The first group includes $\Lambda_{LAF}$, $\Lambda_{AVG}$, and $\Lambda_{MIN}$, which all exhibit similar performance levels. The second group consists of $\Lambda_{MAX}$, while the third group, containing $\Lambda_{RMAX}$ and $\Lambda_{RMIN}$, consistently demonstrates low power across the three relationships considered. Moving forward, the analysis will exclude $\Lambda_{RMAX}$ and $\Lambda_{RMIN}$ due to their consistently low power. Possible reasons for their poor performance will be explored in the next section.

The ranking pattern for each population can be described as follows. 
\begin{itemize}
\item For the Thai population, $\Lambda_{LAF}$ exhibits the highest power in both the parent-child test (83.5\%) and the full-sibling test (78.1\%), followed by $\Lambda_{MIN}$, $\Lambda_{AVG}$, and then $\Lambda_{MAX}$. This result aligns with findings from the Colorado population study in the literature. Notably, $\Lambda_{MIN}$, which was previously suggested to outperform $\Lambda_{MAX}$ at the likelihood ratio distribution level, also demonstrates superior statistical power.
\item In the Singaporean population, the top-performing statistics for the parent-child and full-sibling tests are closely matched, as $\Lambda_{LAF}$, $\Lambda_{AVG}$, and $\Lambda_{MIN}$ produce nearly identical results. $\Lambda_{MAX}$ shows lower power, approximately 6–8\% less than the leading group in the parent-child test at FPR = 0.00002, and about 3\% lower in the full-sibling test at FPR = 0.0002.
\item For the Malaysian subpopulation, we primarily focus on the half-sibling test, as the parent-child and full-sibling tests exhibit relatively low power at the specified FPRs. In this case, $\Lambda_{MIN}$ and $\Lambda_{AVG}$ perform best, with no clear leader for the half-sibling test at $\alpha = 0.002$, followed by $\Lambda_{LAF}$ and $\Lambda_{MAX}$.
\end{itemize}

To facilitate a meaningful comparison, we expanded our analysis by introducing an additional statistic that disregards subpopulation structure. Specifically, we assumed a homogeneous population and designated this statistic as $\Lambda_{CB}$ or LRCB, where "CB" stands for "combine," reflecting the merging of all subpopulations into a single homogeneous group. In other words, $\Lambda_{CB}$ serves as a benchmark for evaluating the performance of statistics both with and without accounting for population substructure.

\begin{figure}[h!]
\includegraphics[scale=0.35]{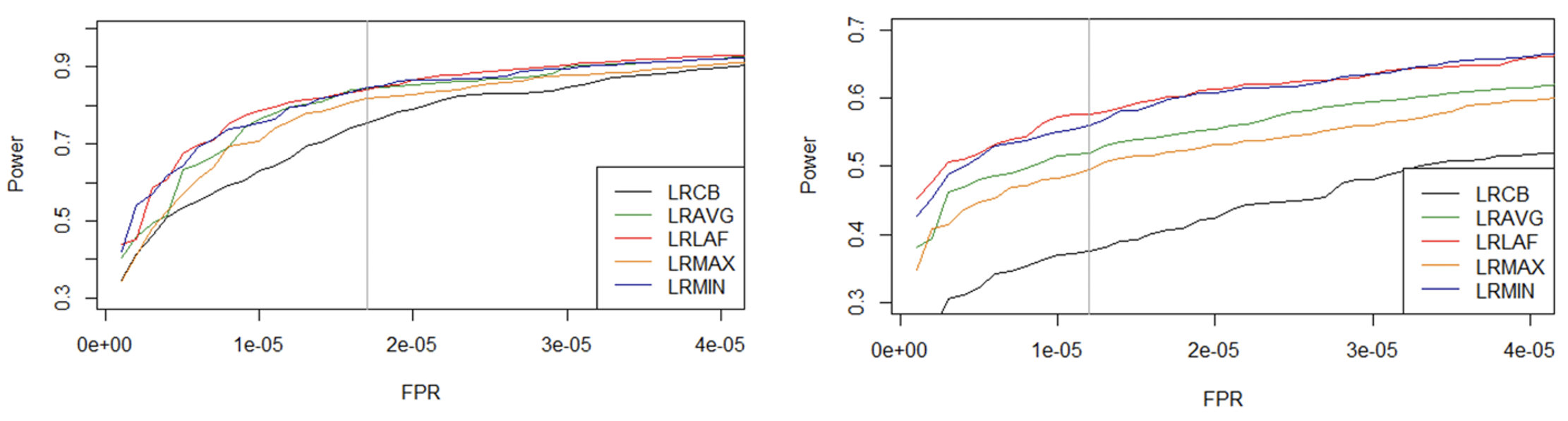}
\caption{\label{fig:power1} Plot of power versus FPR for five different statistics. 
The plot on the left represents parent-child testing, while the one on the right represents full-sibling testing.}
\end{figure}

To evaluate the relative performance ranking of the five statistics—$\Lambda_{LAF}$, $\Lambda_{AVG}$, $\Lambda_{MAX}$, $\Lambda_{MIN}$, and $\Lambda_{CB}$—across a reasonable range of false positive rates (FPRs), power was estimated for significance levels ($\alpha$) ranging from approximately $10^{-6}$ to $4 \times 10^{-5}$. Figure~\ref{fig:power1} (left) shows the power versus false positive rate for these five statistics in parent-child relationship tests. The dashed vertical line represents the estimated false positive rate of 0.000017, as reported in a previous study \cite{kooakachai2019new}. Similarly, Figure~\ref{fig:power1} (right) illustrates the power versus false positive rate for the five statistics in full-sibling relationship tests, with the dashed vertical line indicating an estimated false positive rate of 0.000012, also derived from the same study.

As a result from Figure~\ref{fig:power1}, in the full-sibling test, the ranking pattern of the statistics remains consistent across the considered range of false positive rates, except for $\Lambda_{LAF}$ and $\Lambda_{MIN}$. For false positive rates exceeding $2 \times 10^{-5}$, no clear distinction can be made between $\Lambda_{LAF}$ and $\Lambda_{MIN}$ in terms of performance. A close pattern is observed in the parent-child case, where $\Lambda_{LAF}$, $\Lambda_{AVG}$, and $\Lambda_{MIN}$ perform well with no clear distinction beyond $2 \times 10^{-5}$. Overall, it is clearly evident that the four test statistics incorporating population substructure consistently achieve higher power than $\Lambda_{CB}$.

\subsection{\label{results:powercomp}  Bias towards Subpopulations}

In this subsection, we examined the power for each Thai subpopulation using the eight-step algorithm outlined in Section 2.1. The distinction here is that, when calculating power, we limit the sample to include only pairs from the specific subpopulation under consideration, rather than using the full set of pairs from the alternative distribution. For example, to calculate the parent-child power for the Northeastern subpopulation at an FPR of 0.000017 using the $\Lambda_{MIN}$ statistic, we only select the parent-child pairs from the alternative hypothesis distribution that were generated under the assumption of Northeastern descent. The power is then computed as the proportion of those Northeastern pairs whose $\Lambda_{MIN}$ value exceeds the null threshold corresponding to the FPR of 0.000017. It is important to note that the denominator includes the total number of simulated pairs from the Northeastern subpopulation. This process is repeated for each statistic and each subpopulation.

\begin{figure}[h!]
\centering
\includegraphics[scale=1]{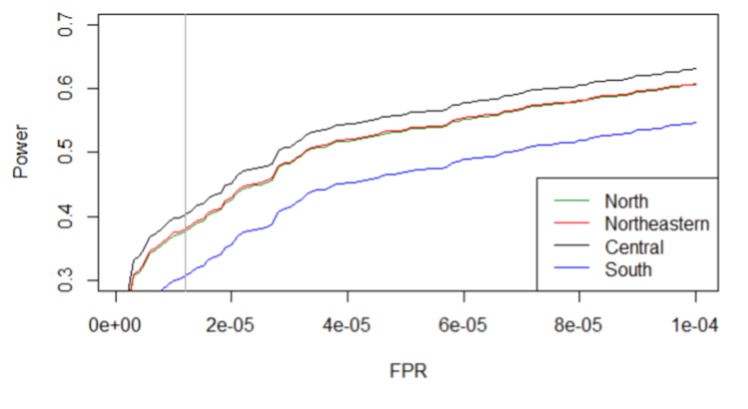}
\caption{\label{fig:power2} Plot of full-sibling power versus FPR for the LRCB statistic, categorized by subpopulation.}
\end{figure}

We created plots showing power versus FPR for each subpopulation. Figure~\ref{fig:power2} presents the plot of full-sibling power versus FPR for the $\Lambda_{CB}$ statistic, broken down by subpopulation. The vertical line indicates the estimated false positive rate of 0.000012. The plot reveals considerable differences in full-sibling power across subpopulations, with the South subpopulation exhibiting the lowest power and the Central subpopulation showing the highest. This suggests that the $\Lambda_{CB}$ statistic, which does not account for subpopulation structure, may not be the best choice, as it leads to biased power variations between subpopulations. To achieve more consistent results in relatedness testing, it is crucial to use a statistic that provides uniform power across all subpopulations.

To identify the best statistic from all available options, the ideal choice would be the one that demonstrates high power while minimizing subpopulation bias. Based on the results from the previous subsection, $\Lambda_{LAF}$ and $\Lambda_{MIN}$ emerge as the top candidates in terms of power, so we will keep them for the next phase of analysis. To make a final decision between $\Lambda_{LAF}$ and $\Lambda_{MIN}$, we will calculate confidence intervals for the difference in power using normal quantiles. Let $\widehat{\beta}_1$ and $\widehat{\beta}_2$ represent the estimated power for the first and second subpopulations, respectively. Then, a $(1-\alpha)100\%$ confidence interval for the difference in power is given by: $$(\widehat{\beta}_1 - \widehat{\beta}_2) \pm z_{\alpha/2}\sqrt{\dfrac{\widehat{\beta}_1(1-\widehat{\beta}_1)}{N_1} + \dfrac{\widehat{\beta}_2(1-\widehat{\beta}_2)}{N_2}}$$
where $z_{\alpha/2}$ denotes the standard normal quantiles at the significance level $\alpha$, and $N_k$ represents the size of subpopulation $k$. It is important to note that this formula is derived from the normal confidence interval, which is valid given the large sample size.

\begin{figure}[h!]
\centering
\includegraphics[scale=0.29]{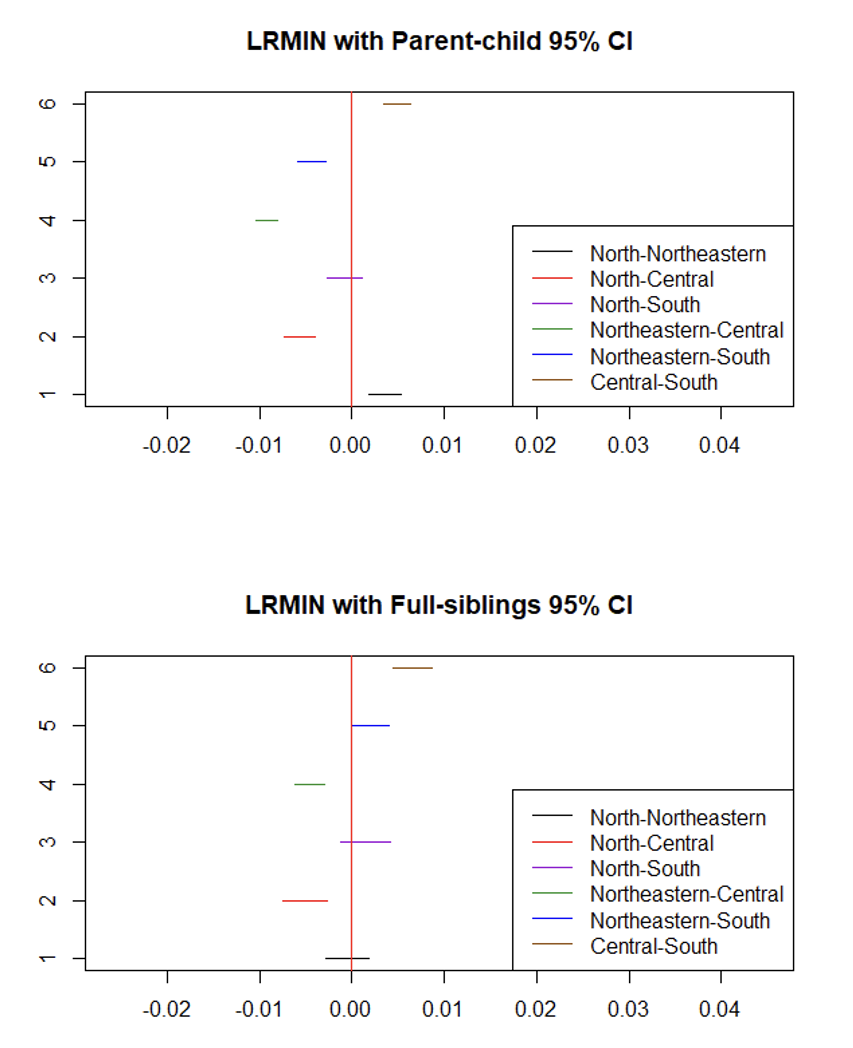}
\caption{\label{fig:power3} Plot of 95\% confidence intervals for the difference in power, broken down by two statistics and two relationships. }
\end{figure}

\begin{figure}[h!]
\centering
\includegraphics[scale=0.29]{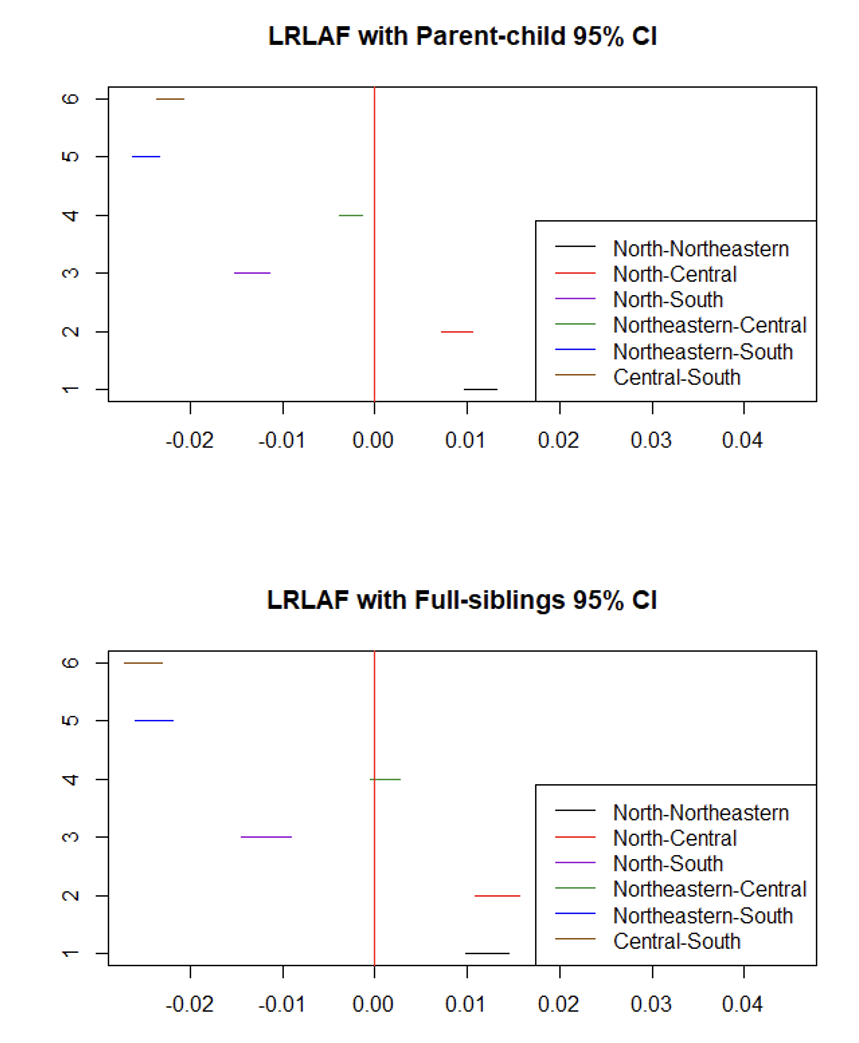}
\caption{\label{fig:power4} Plot of 95\% confidence intervals for the difference in power, broken down by two statistics and two relationships. }
\end{figure}

Figures~\ref{fig:power3} and \ref{fig:power4} illustrate the 95\% confidence intervals for the parent-child and full-sibling tests using $\Lambda_{MIN}$ and $\Lambda_{LAF}$, respectively. The $x$-axis represents the difference in power, while the $y$-axis shows the ordering of pairs of two selected Thai subpopulations out of four. The ideal statistic is one that includes zero within its interval. When the confidence interval encompasses zero, it indicates that zero is a plausible value for the difference in power, suggesting insufficient evidence to claim that the power of the first subpopulation differs from that of the second at the 0.05 significance level. In our study, we prefer the confidence interval to include zero, as this promotes fairness in hypothesis testing for each subpopulation. In both the parent-child and full-sibling tests, we found that $\Lambda_{MIN}$ outperforms the well-established $\Lambda_{LAF}$, as the confidence intervals are closer to zero.

\section{Discussion and Conclusions}

In this study, we introduce several test statistics for hypothesis testing to assess the relationship between two individuals. The available options include $\Lambda_{LAF}$, $\Lambda_{AVG}$, $\Lambda_{MAX}$, $\Lambda_{MIN}$, $\Lambda_{RMAX}$, and $\Lambda_{RMIN}$. Our primary findings clearly indicate that the statistic $\Lambda_{MIN}$ performs exceptionally well in both parent-child and full-sibling tests, achieving power levels comparable to the well-established $\Lambda_{LAF}$. We strongly recommend the practical use of $\Lambda_{MIN}$, especially for Thai subpopulations, for two main reasons. First, $\Lambda_{MIN}$ minimizes bias toward subpopulations, as shown in Figure~\ref{fig:power3}, helping to promote fairness in DNA testing across all subpopulations. Second, it is easier to access relevant datasets, as many standard racial group allele frequencies are readily available in the literature. This allows $\Lambda_{MIN}$ to take advantage of these pre-existing datasets, making it a convenient and practical option. While $\Lambda_{LAF}$ is also a strong choice in terms of power, it requires the use of a local average allele frequency distribution, and in practice, this may necessitate estimating a local average allele frequency from an existing database.

The performance of $\Lambda_{AVG}$ is also comparable to the top two statistics, $\Lambda_{LAF}$ and $\Lambda_{MIN}$, under certain parameter settings, although further investigation is required. As for $\Lambda_{MAX}$, its power is lower than that of the first three statistics, particularly when compared to $\Lambda_{MIN}$. These findings are consistent with the previous study by Ge and Budowle (2012) \cite{ge2012kinship}, which examined results at the likelihood ratio level. We have confirmed this outcome through our analysis of statistical power.

The statistic $\Lambda_{RMAX}$ performed similarly to $\Lambda_{LAF}$ in the Colorado population, but surprisingly, it showed poorer performance in Southeast Asian populations. This highlights the importance of being cautious when conducting studies in this field, as statistical power is highly sensitive to the population under study. As shown here, changing the population can lead to different results for the same test statistic, such as $\Lambda_{RMAX}$. Another notable finding is that $\Lambda_{RMIN}$ performed poorly. This statistic was initially motivated by the strong performance of $\Lambda_{MIN}$, but our results suggest that separately calculating likelihood functions for the null and alternative hypotheses worsens the situation, at least in the three Southeast Asian populations. This is evident from the extremely high null thresholds obtained, as shown in Table~\ref{tab:table3}, which likely makes it much harder to reject the null hypothesis, ultimately leading to low power.

Furthermore, the results in Figure~\ref{fig:power1} emphasize the critical role of accounting for subpopulation structure in genetic analyses, as they reveal a notable decrease in the power of $\Lambda_{CB}$ when substructure is not considered. Additionally, the bias towards subpopulations becomes more apparent when subpopulation structure is not incorporated, as shown in Figure~\ref{fig:power2}. However, the observed bias could be attributed to other factors, such as differences in allelic diversity between subpopulations. For instance, it is well-known that African populations have greater genetic diversity, making them more suitable for familial identification compared to Native American populations. Therefore, it is not surprising that some differences persist.

In comparing power across populations, we find a strong consistency among Southeast Asian populations, with $\Lambda_{LAF}$, $\Lambda_{MIN}$, and $\Lambda_{AVG}$ emerging as the top performers. In contrast, $\Lambda_{MAX}$, $\Lambda_{RMAX}$, and $\Lambda_{RMIN}$ consistently exhibit lower power compared to this leading group. The lower power observed for $\Lambda_{MAX}$ aligns with similar findings in studies of North American populations. Additionally, the Malaysian population, which only used 9 loci in this study, showed lower power compared to the Thai and Singaporean populations, which used 15 loci. This result is expected, as a higher number of loci typically leads to greater power due to increased variation available for detection. To boost the power of the Malaysian population to match that of the Thai or Singaporean populations, increasing the number of loci would be a recommended approach.

In concluding this paper, we propose several avenues for future research. First, while our hypothesis tests primarily concentrated on parent-child, full-sibling, and half-sibling relationships, exploring more intricate familial relationships, such as first cousins, second cousins, and double first cousins, would be a valuable next step. Investigating how statistical power varies as the complexity of relationships increases could provide valuable insights into relatedness testing. However, care must be taken when selecting the false positive rates to appropriately differentiate between the performance of each test statistic. Second, while $\Lambda_{LAF}$ and $\Lambda_{MIN}$ have proven to be effective, there may be other statistics that outperform them in terms of power. Although both of these statistics are robust, it remains unclear whether they are the most powerful available. At present, there is no established theory that identifies the most universally powerful statistic in this context. One potential direction for future research could be to employ machine learning or classification methods to first identify the subpopulation to which a DNA profile belongs, followed by applying likelihood ratio specific to that subpopulation. This approach may lead to improved performance in relatedness testing, especially when subpopulation structure is considered more explicitly. Finally, our analysis was based on a Thai and Southeast Asian population with regional substructure. A valuable extension of this research would be to replicate these tests in different populations with their own unique substructure. By comparing results across a variety of populations, we could better understand the limitations of our findings, and explore how subpopulation structure influences relatedness testing in diverse genetic contexts. 

\section*{Acknowledgement}
The authors would like to express their heartfelt gratitude to the editor, associate editor and referees for their insights and suggestions. This research project is supported by Ratchadapiseksompotch Fund Chulalongkorn University.

\end{document}